\documentclass[12pt]{article}
\usepackage{amssymb}
\def\g{{\gamma}}

\def\l{{\lambda}}

\def\X{{\cal X}}
\def\J{{\cal J}}

\def\te#1{{\widetilde{#1}}}

\def\nn{ \nonumber }
\def\bq{ \begin{equation} }
\def\eq{ \end{equation} }
\def\ben{ \begin{eqnarray} }
\def\en{ \end{eqnarray} }
\def\frac#1#2{{#1\over #2}}
\def\dfrac#1#2{{\displaystyle{#1\over#2}}}

\newtheorem{prop}{Proposition}

\begin{document}
\title{On integrable deformations of the spherical top.}
\author{
 A.V. Tsiganov\\
{\small\it
 Department of Mathematical and Computational Physics,
 Institute of Physics,}\\
{\small\it St.Petersburg University, 198 904,  St.Petersburg,
Russia,}\\ {\small\it e-mail: tsiganov@mph.phys.spbu.ru} }
\date{}
\maketitle

\vskip0.5cm
{\small  The motion on the sphere $S^2$ in  potential $V=
(x_1x_2x_3)^{-2/3}$ is considered. The Lax representation and the
linearisation procedure for this two-dimensional integrable system
are discussed.
\vskip1truecm}
\vfill
\newpage

\section{Description of the model}
The system under consideration is a special case of the following
mechanical system in the nine-dimensional space ${\mathbb R}^9$
\bq
F^\ast(t)\,\dfrac{d^2\,{F}(t)}{d\,t^2}+ |\det
F(t)|^{1-\gamma}\,G=0\,.\label{syseq1}
\eq
Here $F(t)$ and $G$ are $3\times 3$ matrices, $F^\ast$ means
transpose matrix and $\gamma$ is an polytropic index. The components
$F_{jk}$ of the matrix $F$ are coordinates on the configuration space
${\mathbb R}^9$. These equations of motion have been studied many
authors, see \cite{ov56,dy68,al70,bog80}.

According to \cite{ov56}, we shall consider symmetric constant
matrices $G=G^\ast$ only. In this case, by using canonical
transformation of variables $F'=UFU^\ast$, we can reduce the constant
matrix $G$ to the diagonal matrix with the following diagonal
elements $\pm 1$ or $0$. Moreover, with physical point of view  we
can put $G=I$ without loss of generality \cite{ov56}.

The Newton equations (\ref{syseq1}) arise in solution of the
hydrodynamical equations representing dynamics of a cloud of
compressible gas expanding freely in an otherwise empty space. This
model has a rich history related with the names of Dirichlet,
Dedekind and Riemann. For an extensive discussion of the model we
refer to \cite{bog80}, a book which should be taken as a general
reference on the subject.

At $G=G^\ast$ the invariance of the problem under rotation and under
internal motion of the gas leads to conservation of the angular
momentum and operator of vorticity
\[{\bf J}=F(t)\,\dot{F}^\ast(t)-\dot{F}(t)\,F^\ast(t)\,,\qquad
{\bf K}=F^\ast(t)\,\dot{F}(t)-\dot{F}^\ast(t)\, F(t)\,.\] Each of
$\bf J$, $\bf K$ is an antisymmetric matrix with $3$ independent
components. Thus, equations (\ref{syseq1}) possess an enlarged
symmetry group $SO(4)\simeq SO(3)\times SO(3)$. There is also a
discrete symmetry, which allows the vorticity and the angular
momentum to be interchanged. This discrete symmetry is identical with
the duality principle of Dedekind \cite{dy68}.

For the perfect monatomic gas, at $\gamma=5/3$ the system of
equations (\ref{syseq1}) possesses one more integral of motion
\cite{al70}
\bq
{\bf r}^2={\rm tr}\,\Bigl(\,F^\ast(t)\,F(t)\,\Bigr)\,, \qquad{\rm
at}\quad\g=\dfrac{5}{3}\,. \label{fint}
\eq
The qualitative behavior of solutions at the different values of
the adiabatic index $\g$ and some partial cases of motion are
discussed in \cite{bog80}.

It is easy to separate the diagonal and non-diagonal components of
equations (\ref{syseq1}) \cite{ov56,dy68,bog80}. The non-diagonal
components give six kinematical equations involving only the inertial
properties of the gas-cloud but not the pressure force. The diagonal
components give three dynamical equations determining the rate of
expansion of the gas-cloud under the influence of the pressure force.

Below at $\g=5/3$ we shall consider a free expansion of an
ellipsoidal gas cloud with the fixed orientation, having zero
angular momentum and zero vorticity ${\bf K}={\bf J}=0$. In this
case matrix $F(t)$ is diagonal for all $t$
\[F(t)=diag(\,x_1,\,x_2,\,x_3\,)\,(t)\]
and the constant matrix $G=I$ is equal to unit. The corresponding
three equations of motion are given by
\bq
x_1\,\ddot{x}_1=x_2\,\ddot{x}_2=x_3\,\ddot{x}_3=
\dfrac{const}{(x_1\,x_2\,x_3)^{2/3}}\,, \qquad{\rm
at}\quad\g=\dfrac{5}{3}\,. \label{eq}
\eq
Additional integral of motion (\ref{fint}) \cite{al70} is equal to
the radius of sphere ${\bf r}^2=\sum x_k^2$. By using this
integral our system on ${\mathbb R}^3$  may be reduced to the
system on the sphere $S^2$. At $\g=5/3$ the third independent
integral of motion was founded in \cite{gaf96}.

The reduced system has the configuration space diffeomorphic to
the Euclidean motion group $E(3)=SO(3)\times {\mathbb R}^3$
\cite{arn89}. It allows to identify the phase space of this system
on $T^*S^2$ with the cotangent bundle $T^*E(3)$. The kinetic
energy is a left-invariant Riemannian metric on $E(3)$. It is
determined by some quadratic form on the dual space $e^*(3)$ of
the Lie algebra $e(3)$ \cite{arn89,rs87}.

By using the Killing form the dual space $e^*(3)$ may be
identified with algebra $e(3)=so(3) \oplus {\mathbb R}^3$, the
semi-direct sum of $so(3)$ and the abelian space ${\mathbb R}^3$.
Let two vectors $J\in so(3)\simeq {\mathbb R}^3$ and $x\in
{\mathbb R}^3$ be coordinates in the dual space $e^*(3)$ equipped
with a natural Lie-Poisson brackets
\ben
&&\bigl\{ J_i\,,J_j\,\bigr\}= \varepsilon_{ijk}\,J_k\,, \qquad
\bigl\{ J_i\,,x_j\,\bigr\}= \varepsilon_{ijk}\,x_k\,,\nn\\
&&\bigl\{ x_i\,,x_j\,\bigr\}= 0\,,\qquad i,j,k=1,2,3,. \label{e3}
\en
Here $\varepsilon_{ijk}$ is the standard totally skew-symmetric
tensor. The generic coadjoint orbits of $E(3)$ in $e^*(3)$ are
four dimensional symplectic leaves specified by the two Casimir
elements
\bq
{\bf C}_1=(x,x)=x_i x_i;\qquad {\bf C_2}=(J,x)=J_i x_i\,.
\label{caz}
\eq
Here $(x,y)$ means inner product in ${\mathbb R}^3$. Thus, the
dual space $e^*(3)$ decomposes into the coadjoint orbits
\bq
{\cal O}_{c_1,\,c_2}=\Bigl\{ \{J,x\}\in{\mathbb R}^6\,:~{\bf
C}_1=c_1\,,~{\bf C}_2=c_2\Bigr\}\, \label{orb}
\eq
which are invariant with respect to the usual Euler-Poisson
equations in $e^*(3)$ \cite{arn89,rs87}.

Let us introduce a complex analog of the Lie algebra $e(3)$, as a
semi-direct sum of $so(3,{\mathbb C})$ and the complex space
${\mathbb C}^3$. This algebra $e(3,{\mathbb C})=so(3,{\mathbb C})
\oplus {\mathbb C}^3$ is equipped with the same Lie-Poisson
brackets (\ref{e3}) and the Casimir operators (\ref{caz}). In
contrast with the usual $e(3)$ algebra, it allows us to consider
non-trivial representations at zero value $c_1=0$ of the first
Casimir operator ${\bf C}_1$.

The condition $c_1=0$ has no obvious physical or geometric
meaning. Of course, we can not consider real sphere of zero
radius, but with mathematical point of view $c_1$ is an arbitrary
value of the Casimir element. Note, that "non-physical"
representations of the algebra $sl(2)$ with the zero spin $s=0$
are useful in physics \cite{fk94,ts98p} as well.

\begin{prop}
At zero value $c_1=0$ of the first Casimir operator ${\bf C}_1$
(\ref{caz}) the following transformation in $so^*(3,{\mathbb
C})\subset e^*(3,{\mathbb C})$
\bq
J\to \te{J}=J+\dfrac{i\,a}{(x_1\,x_2\,x_3)^{1/3}}\,x\,, \quad
a\in{\mathbb C} \label{auto}
\eq
is an outer automorphism of the representation of $e(3,{\mathbb
C})$.
\end{prop}

By using embedding $e(3)\subset e(3,{\mathbb C})$ let us consider
known integrable tops on this complex algebra $e^*(3,{\mathbb C})$.
Applying transformation (\ref{auto}) one can get integrable
deformations of these top  on the one-parameter set of orbits ${\cal
O}_1$ ($c_1=0, c_2=const$). Sometimes outer automorphism (\ref{auto})
allows us to get much more.

As an example, let us consider spherical top with the standard
Hamilton function
\bq
H=(\te{J},\te{J})=\te{J}_1^2+\te{J}_2^2+\te{J}_3^2\label{sph}
\eq
and with the
non-standard second integral of motion
\bq K=\te{J}_1\,\te{J}_2\,\te{J}_3 \label{spk}\eq
defined on the subalgebra $so(3)$ only. Of course, substituting
vector $\te{J}$ (\ref{auto}) one gets integrable deformation of this
symmetric top  at $c_1=0$. However, we can prove the following

\begin{prop}
Outer automorphism (\ref{auto}) maps Hamiltonian (\ref{sph}) of the
spherical top into the following Hamiltonian
\bq
H=\sum_{k=1}^3 J_k^2+2ia\,\dfrac{c_2}{(x_1\,x_2\,x_3)^{1/3}}
-a^2\,\dfrac{c_1}{(x_1\,x_2\,x_3)^{2/3}}\,. \label{ham}
\eq
The proposed deformation of the spherical top  is completely
integrable on the both one-parameter sets of orbits
\[{\cal O}_1:\quad(c_1=0,~c_2=\mbox{const})\qquad\mbox{\it and}\qquad  {\cal
O}_2:\quad(c_1=\mbox{const},~c_2=0)\,.\] Moreover, to consider the
second orbits we can return to the usual real Lie algebra $e^*(3)$.
\end{prop}
At $c_1=0$ the second integral is the image of known integral $K$
(\ref{spk})
\ben
K&=&J_1\,J_2\,J_3-a^2\,
\left(\,\dfrac{J_1}{x_1}+\dfrac{J_2}{x_2}+\dfrac{J_3}{x_3}\right)
\,(x_1\,x_2\,x_3)^{1/3}\nn\\ \label{spkk}\\
&+&\dfrac{2\,i\,a}{(x_1\,x_2\,x_3)^{1/3}}\,\left(J_1\,J_2\,x_3
+J_1\,x_2\,J_3+x_1\,J_2\,J_3 \right)-i\,a^3\,.\nn
\en
At $c_2=0$ the Hamilton function (\ref{ham}) is in involution with
the following second integral of motion
\bq
K=J_1\,J_2\,J_3+
a^2\,\left(\,\dfrac{J_1}{x_1}+\dfrac{J_2}{x_2}+\dfrac{J_3}{x_3}\right)
\,(x_1\,x_2\,x_3)^{1/3}\,.\label{int}
\eq
In contrast with (\ref{spkk}), here we removed imaginary terms and
changed the sign before the rest potential term. We do not know
origin of such additional transformation as yet.

In the natural variables
\[y=\dfrac{x}{(x_1\,x_2\,x_3)^{1/3}}\,,\qquad y_1\,y_2\,y_3=1\]
transformation (\ref{auto}) becomes a shift $\te{J}=J+i\,a\,y$ and
the integrals of motion (\ref{ham}) and (\ref{int}) are given by
\ben
H&=&J_1^2+J_2^2+J_3^2-a^2\,(y_1^2+y_2^2+y_3^2)\,,\nn\\
\label{ints}\\
K&=&J_1\,J_2\,J_3+a^2\,\left(\,\dfrac{J_1}{y_1}+\dfrac{J_2}{y_2}+
\dfrac{J_3}{y_3}\right)\,.\nn
\en
The Euler-Poisson equations on $e^*(3)$ generated by (\ref{ham})
are given by
\ben
\dfrac{d}{dt}\,{J}&=\dfrac{2\,a^2}{3}\,(y,y)\,&y\times y^{-1}\,,
\qquad y^{-1}=(y_1^{-1},y_2^{-1},y_3^{-1})\,, \nn\\
\label{euleq}\\ \dfrac{d}{dt}\,y&=-\dfrac{2}{3}\,(y,y)\,&J\times
y^{-1}\,, \qquad (y,y)=y_1^2+y_2^2+y_3^2\,, \nn
\en
where $x\times y$ means standard vector product in ${\mathbb R}^3$.
Thus, we rewrite initial very symmetric equations of motion
(\ref{eq}) defined on the configuration space ${\mathbb R}^3$ as the
Euler-Poisson equations (\ref{euleq}) defined on the phase space
$e^*(3)$.

\section{Lax representation}
The main purpose of this note is to rewrite equations of motion
(\ref{euleq}) in the Lax form
\bq
\dfrac{d}{dt}\,L=[L,M]\,. \label{laxdef}
\eq

Let us briefly recall construction of the Lax pair for the Neumann
system. The Neumann system is an integrable system on the sphere
with quadratic potential (see (\ref{ints})). Its phase space may
be modelled on the dual space $e^*(3)$ at $c_2=0$. The
corresponding Euler-Poisson equations are equal to
\bq
\dfrac{d}{dt}\,{J}=x\times z\,, \qquad \dfrac{d}{dt}\,x=-J\times
x\,,\qquad z=-{\rm diag}(a_1,a_2,a_3)\,x\,.\label{euleq1}
\eq
Here $a_j$ be arbitrary parameters.

The Neumann system possesses the necessary number of the quadratic
integrals of motion. Nevertheless, the Lax pair can not be
constructed in framework of the algebra $e(3)=so(3) \oplus
{\mathbb R}^3$. Namely, for the Neumann system and some others
system, we have to use the Cartan-type decomposition of the Lie
algebra $gl(3,{\mathbb R})=so(3)+Symm(3)$ \cite{rs87}.

Let us introduce antisymmetric matrix of angular momentum $\J\in
so(3)$ and symmetric matrix of coordinates $\X\in Symm(3)$
\ben
&\J\in so(3)\simeq{\mathbb R}^3:\qquad
&\J_{ij}=\varepsilon_{ijk}\J_k\,,\nn\\ \nn\\ &\X\in Symm(3):\qquad
&\X_{ij}=x_i\,x_j\,.\nn
\en
Then the Lax representation for the Neumann system are given by
\bq
L={\rm diag}(a_1,a_2,a_3)\,\lambda+\J+\lambda^{-1}\,\X\,,
\qquad M=-\lambda^{-1}\,\X\,.\label{laxneu}
\eq
Let us present these Lax matrices explicitly
\[
L=\left(\begin{array}{ccc}a_1\l&0&0\\0&a_2\l&0\\0&0&a_3\l\end{array}\right)\,+
\left(\begin{array}{ccc}0&-J_3&J_2\\ J_3&0&-J_1\\
-J_2&J_1&0\end{array}\right)\,+\dfrac{1}{\l}\,
\left(\begin{array}{ccc}x_1^2&x_1\,x_2&x_1\,x_3\\
 x_1\,x_2&x_2^2&x_2\,x_3\\x_1\,x_3&x_2\,x_3&x_3^2
\end{array}\right)\,,\]
\[M=-\dfrac{x_1\,x_2\,x_3}{\l}\,
\left(\begin{array}{ccc}
\\
\dfrac{x_1}{x_2\,x_3}&x_3^{-1}&x_2^{-1}\\
x_3^{-1}&\dfrac{x_2}{x_1\,x_3}&x_1^{-1}\\
x_2^{-1}&x_1^{-1}&\dfrac{x_3}{x_1\,x_2}\\
\\
\end{array}\right)\,.
\]

Now let us turn to the deformation of the completely symmetric top
(\ref{ham}). The Lie algebras ${\mathbb R}^3$ with vector product
and $so(3)$ with usual commutator may be identified by using the
Lie algebras isomorphism
\[x=\left(\begin{array}{c}x_1\\ x_2\\ x_3\end{array}\right)\in{\mathbb R}^3
\longrightarrow \X=\left(\begin{array}{ccc}0&x_3&-x_2\\
-x_3&0&x_1\\x_2&-x_1&0
\end{array}\right)\in so(3)\,.
\]
In (\ref{auto}) the element $J\in so(3)$ has been added with the
vector $x\in {\mathbb R}^3$. Thus, defining outer automorphism
(\ref{auto}), we implicitly used this property of the three
dimensional Euclidean space. Below we shall use the same property
to construct the Lax representation.

The main recipe is to rearrange items in decomposition
$gl(3,{\mathbb R})=so(3)+Symm(3)$. Let us introduce symmetric
matrix of angular momentum $\J\in Symm(3)$ and antisymmetric
matrix of coordinates $\X\in so(3)$
\ben
&\J\in Symm(3):\qquad
&\J_{ij}=|\,\varepsilon_{ijk}\,|\,J_k\,,\nn\\ \nn\\ &\X\in
so(3)\simeq{\mathbb R}^3:\qquad &\X_{ij}=\varepsilon_{ijk}\,
y_k\,,\nn
\en
where $|\,\varepsilon_{ijk}\,|$ means absolute value of
$\varepsilon_{ijk}$.
\begin{prop}
At $c_2=0$ the equations of motion (\ref{euleq}) on the sphere
$S^2$ generated by the Hamilton function (\ref{ints}) can be
written in the Lax form (\ref{laxdef}) with the following matrices
\bq
L=\lambda\,I+\J+a\,\X\,, \qquad
M_{ij}=\dfrac{2a}{3}\,|\,\varepsilon_{ijk}\,|\,x_k^{-1}
\,.\label{laxgaf}
\eq
More explicitly, the first matrix is
\[
L=\left(\begin{array}{ccc}\l&0&0\\0&\l&0\\0&0&\l\end{array}\right)\,+
\left(\begin{array}{ccc}0&J_3&J_2\\ J_3&0&J_1\\
J_2&J_1&0\end{array}\right)\,+\,a\,
\left(\begin{array}{ccc}0&y_3&-y_2\\ -y_3&0&y_1\\y_2&-y_1&0
\end{array}\right)\,,\]
and the second matrix is given by
\[
M=\dfrac{2\,a\,c_1^{1/2}}{3\,\sqrt{\,(y,y)\,}}\,
\left(\begin{array}{ccc}0&y^{-1}_3&y^{-1}_2\\
y^{-1}_3&0&y^{-1}_1\\ y^{-1}_2&y^{-1}_1&0\end{array}\right)
\]
The spectral invariants of $L(\l)$ give rise to both integrals of
motion in involution (\ref{ints})
\[\det L(\l)=\l^3-H\l+2K\,.\]
\end{prop}
The proposed Lax matrix $L(\l)$ has a trivial dependence on spectral
parameter $\l$, which is similar to the Lax matrix for the Kowalewski
top due by Perelomov \cite{rs87}. Therefore, we can not construct a
suitable spectral curve and can not directly integrate equations of
motion. Recall, in \cite{rs87} the Perelomov matrices have been
embedded into the general Lax matrices with the spectral parameter.
It forces us to consider the Lax representation (\ref{laxgaf}) as a
first attempt to build an adequate Lax pair.  We believe the desire
Lax pair explains the peculiar geometry and the origin of
integrability of the considered motion on the sphere $S^2$.

\section{Linearisation procedure}
In closing this note we briefly discuss results obtained in
\cite{gaf98} within the modern theory of linearisation of the
two-dimensional integrable systems \cite{avm89,bvm87,van92}. The
proposed in \cite{gaf98} procedure of integration has an unusual
form, which is closely related with the concrete system of equations.
This construction is technically closed to the Chaplygin approach
\cite{ch48} to the Kirchhoff equations at $c_2=0$. On the other hand,
modern theory of linearisation allows us to consider different
integrable systems such as the Neumann problem, Henon-Hailes system,
Toda lattice, Kowalewski top, Goryachev-Chaplygin top and many other
\cite{avm89,bvm87,van92}.

However, in this common powerful method it is necessary to rewrite
equations of motion at some suitable variables. These variables have
to satisfy the special conditions \cite{avm89,bvm87,van92}. As an
example, to integrate the Toda lattice we have to introduce so-called
Flaschka variables. For other system such variables  may be
introduced by using the  Kowalewski-Painlev\'{e} analysis or the
algebro-geometric tools. Nevertheless, if we might have introduced
such variables, the Adler and van Moerbeke methods
\cite{avm89,bvm87,van92} enable us to integrate a given mechanical
system.

The aim of this Section is to introduce an analog of the Flaschka
variables for the deformations of the spherical top. At these
variables we may directly apply  the Adler and van Moerbeke methods
to a given integrable system. These results will be published at the
consequent publications.

The three body Toda lattice is the Hamiltonian system defined as
\[H=\dfrac12\sum_{j=1}^3p_j^2+e^{q_1-q_2}+e^{q_2-q_3}+e^{q_3-q_1}\,.\]
Here $\{p_j,~q_j\}_{j=1}^3$ be pairs of the canonical physical
variables. According to \cite{avm89,bvm87,van92}, in the Flaschka
variables
\ben
&&z_1=e^{q_1-q_2}\,,\quad z_2=e^{q_2-q_3}\,,\quad
z_3=e^{q_3-q_1}\,,\qquad z_1\,z_2\,z_3=1\nn\\ &&z_4=-p_1\,,\qquad
z_5=-p_2\,,\qquad z_6=-p_3 \nn
\en
the corresponding equation of motion have the following form
\ben
&\dfrac{d}{dt}\,z_1=z_1\,(z_5-z_4)\,,\qquad&\dfrac{d}{dt}\,z_4=z_1-z_3\,,\nn\\
&\dfrac{d}{dt}\,z_2=z_2\,(z_6-z_5)\,,\qquad&\dfrac{d}{dt}\,z_5=z_2-z_1\,,
\label{todaeq}\\
&\dfrac{d}{dt}\,z_3=z_3\,(z_4-z_6)\,,\qquad&\dfrac{d}{dt}\,z_6=z_3-z_2\,,\nn
\en
The Toda flow has the following four constants of motion
\ben
&&Z_1=z_1\,z_2\,z_3=1\,,\nn\\ &&Z_2=z_4+z_5+z_6=d_1=0\,,\nn\\
&&Z_3=\dfrac12\,(z_4^2+z_5^2+z_6^2)+z_1+z_2+z_3=a_1\,,\label{mantoda}\\
&&Z_4=z_4\,z_5\,z_6-z_1\,z_6-z_2\,z_4-z_3\,z_5=b_1\,.\nn
\en
At $d_1\neq 0$ the variable $Q=q_1+q_2+q_3$ can not be restored from
variables $\{z\}_{j=1}^6$. Really, we have to add to variables
$\{z_j\}$ some other variables with the trivial dynamics
\cite{avm89}. Below we shall introduce analog of the Flaschka
variables for the integrable deformations of the spherical top.

There is a discrete permutation group acting on the vectors $q$ and
$p$ simultaneously
\bq
q\longmapsto {\cal D}\,q\,,\qquad p\longmapsto {\cal
D}\,p\,,\qquad {\cal D}=
\left(\begin{array}{ccc}0&1&0\\0&0&1\\1&0&0\end{array}\right)\,,\qquad
{\cal D}^3=1.\label{auttoda}
\eq
According this point symmetry, the invariant manifold defined by
(\ref{mantoda}) has the third order automorphism  given by
\bq
(\,z_1,\,z_2,\,z_3,\,z_4,\,z_5,\,z_6\,)\longmapsto
(\,z_2,\,z_3,\,z_1,\,z_5,\,z_6,\,z_4\,)\,. \label{auttoda1}
\eq
This automorphism simplifies the  Adler and van Moerbeke analysis
\cite{avm89,bvm87}, which applied to this system,
gives linearisation to the Toda flow.

Recall, the Kowalewski-Painlev\'{e} analysis enables us to integrate
equations of motion for the Goryachev-Chaplygin top
\cite{bvm87,van92} as well. It is another integrable system on
$e^*(3)$ at $c_2=0$. In this case we have to introduce some
seven-dimensional system with the five constants of motion. Then this
seven-dimensional system may be reduced to the Toda system
\cite{bvm87}. The similar relations between the Toda flow and the
integrable system on $e^*(3)$ is discussed in \cite{kuzts89}. Now we
want to compare the Toda flow with another integrable system on
$e^*(3)$.

Let us turn to the deformation of the completely symmetric top
(\ref{ham}). In \cite{gaf96}, the following transformation of the
independent time variable was proposed
\bq
t\to u\,:\qquad
\dfrac{d}{du}=\dfrac{4}{3}\,(y,y)\,\dfrac{d}{dt}\,,
\label{time}
\eq
because of the "weak" Kowalewski-Painlev\'{e} criterion. In the
new time variable the initial Euler-Poisson equations
(\ref{euleq}) are
\bq
\dfrac{d}{du}\,{J}= \dfrac{a^2}2\,y\times y^{-1}\,, \qquad
\dfrac{d}{du}\,y=-\dfrac{1}2\,J\times y^{-1}\,. \label{euleq2}
\eq
Namely these equations were integrated in hyperelliptic
quadratures in \cite{gaf98}.

If we want to compare integrable system on the sphere $S^2$ with
the Toda system, note that equations (\ref{todaeq}) are invariant
by
\bq
z_j\longmapsto \dfrac{1}{z_j}\,,\qquad z_{j+3}\longmapsto
-z_{j+3}\,,\qquad j=1,2,3\,. \nn
\eq
The first equation in
(\ref{euleq2}) has the similar property by
\[y_j\longmapsto \dfrac{1}{y_j}\,,\qquad J_j\longmapsto -J_j\,,
\qquad j=1,2,3\,.\]
Using this observations we introduce analog of the Flaschka variables
\ben
&&s_1= y_1^{-2}\,,~\qquad s_2= y_3^{-2}\,,~~\qquad s_3=
y_2^{-2}\,,
\qquad s_1\,s_2\,s_3=1\,,\nn\\ &&s_6= y_1\,J_1\,,\qquad s_4=
y_3\,J_3\,,\qquad s_5= y_2\,J_2\,.\nn
\en
satisfying the following equations
\ben
&\dfrac{d}{du}\,s_1=s_1(s_5-s_4)\,,\qquad&
\dfrac{d}{du}\,s_4=\dfrac{a^2}2\,(s_1-s_3)+\dfrac{s_4}2\,(s_5-s_6)\,,\nn\\
&\dfrac{d}{du}\,s_2=s_2(s_6-s_5)\,,\qquad&
\dfrac{d}{du}\,s_5=\dfrac{a^2}2\,(s_2-s_1)+\dfrac{s_5}2\,(s_6-s_5)\,,\label{seq}\\
&\dfrac{d}{du}\,s_3=s_3(s_4-s_6)\,,\qquad&
\dfrac{d}{du}\,s_6=\dfrac{a^2}2\,(s_3-s_2)+\dfrac{s_6}2\,(s_4-s_5)\,.\nn
\en
The first column of equations is completely coincided with the
corresponding Toda equations (\ref{todaeq}). The second columns
differ on the at most quadratic polynomials. Thus, for these
equations we can directly apply the linearisation procedure due by
Adler and van Moerbeke \cite{avm89,bvm87}. The freedom in definition
of the variables $s_j$ may be used to make the Laurent solutions a
bit easier.

The constants of motion  for the flow (\ref{seq}) are given by
\ben
&&S_1=s_1\,s_2\,s_3=1\,,\nn\\ &&S_2=s_4+s_5+s_6=c_2=0\,,\nn\\
&&S_3=(\,s_4^2\,s_2+s_5^2\,s_3+s_6^2\,s_1\,)
-a^2\,(s_1\,s_2+s_1\,s_3+s_2\,s_3)=a_1\,,\label{mangaf}\\
&&S_4=s_4\,s_5\,s_6+a^2\,(s_1\,s_6+s_2\,s_4+s_3\,s_5)=b_1\,,\nn
\en
The three constants $S_1,~S_2$ and $S_4$ coincides with the
corresponding Toda constants. The Hamiltonian $S_3$ is a qubic
polynomials now.

There is the permutation invariance similar to the Toda lattice
(\ref{auttoda}). Thus, the invariant manifold defined by
(\ref{mangaf}) possesses the third order  automorphism
(\ref{auttoda1}), which simplifies the linearisation procedure.

It may of course be quite difficult to find variables similar to
$\{z_j\}$ or $\{s_j\}$ associated with a given mechanical system.
These variables have to satisfy some special conditions
\cite{avm89,van92}. For instance, the corresponding equations of motion
have to include at most second order polynomials only,see
(\ref{todaeq},\ref{seq}). However, if we can introduce such
variables, the Adler and van Moerbeke methods
\cite{avm89,bvm87,van92} enable us to integrate a given mechanical
system.

Thus, for a motion on sphere (\ref{ints}) one has to embed the
affine invariant surface defined by (\ref{mangaf}) into the
projective space, whose closure is a principally polarised Abelian
surface. It enables one to define the system to linearising
variables. Then we have to prove that the vector field
corresponding to $S_4$ (\ref{mangaf}) gives the highest flow with
respect to the same hyperelliptic curve of genus two. It will
complete the linearisation of the integrable deformation of the
spherical top. Of course, this general machinery leads to the
particular results obtained in \cite{gaf98}.

\section{Conclusion}
In the algebro-geometric approach due to Adler and van Moerbeke
\cite{avm89,bvm87,van92}, the algebraic curve may be constructed
without any Lax pair representation. For the considered motion on
sphere (\ref{ham}), by substituting the Laurent solutions in the
invariants (\ref{mangaf}) one gets the hyperelliptic curve
\cite{gaf98}. Starting with this curve and the linearising
variables \cite{gaf98} the $2\times 2$ Lax pair may be obtained
(see \cite{van92} for a review).

In this note we tried to construct the Lax pair in framework of
the group-theoretical approach to integrable system
\cite{rs87,brs89}. Applying the method of finite-band integration
to the adequate Lax matrix, we hope to get solutions, which may be
more simpler than the original formulae \cite{gaf98}, as for the
Kowalewski top \cite{brs89}.

Further properties of the integrable deformation of the completely
symmetric top, like action-angles variables, Poisson structures of
the seven-dimensional system and separation of variables are under
study. The results and the more detailed geometric description
will be published elsewhere.

\section{Acknowledgement}
Author is grateful to I.V.Komarov who attracted his attention to
the problem discussed. This work was partially supported by RFBR
and GRACENAS grants.

\end{document}